# Very Energetic Gamma-Rays from Microquasars and Binary Pulsars

I.F. Mirabel[1]

Compact astrophysical objects produce some of the highest energy light in the universe. The challenge is to determine what mechanism produces these photons.

A new window on the universe is presently being opened by ground-based telescopes that survey the sky by detecting very high energy (VHE) photons, which have energies greater than 100 gigaelectron volts (GeV). Because of their high sensitivity, and high angular and energy resolution, these telescopes are revealing and identifying a plethora of new extragalactic and galactic sources of VHE radiation. The Galactic Center, supernovae remnants, pulsar-wind nebulae, and a new class of binary stars called gamma-ray binaries have all been identified as VHE sources in the Milky Way. In this issue, Albert et al. (1) confirm the identification of LSI +61 303 as the third gamma-ray stellar binary known so far, and report a time variability in the signal that may point to a mechanism for the VHE emission.

A microquasar-jet (2) model (see the figure, left panel) has been proposed to account for the VHE emission from another gamma-ray binary, LS 5039 (3). For LSI +61 303, Albert et al. favor a mechanism, called inverse Compton scattering, by which relativistic particles collide with stellar and/or synchrotron photons and boost their energies to the VHE range (4,5). VHE photons have also been detected from blazars, namely, active galactic nuclei (AGN) whose jets are closely aligned with our line of sight. Because the particle energy in microquasar jets is comparable to that of particles in AGN jets (2), it is expected that microquasars with jets pointing to the Earth may appear as scale-down versions of blazars, which have been named microblazars (6). This idea has been strengthened by recent observations showing that the kinetic power of microquasar jets may be larger than $10^{39}$ ergs s$^{-1}$, which is larger than the radiated power (7). Furthermore, microquasar jets trigger shocks where electrons are accelerated up to TeV energies (8), providing the necessary conditions for VHE emission.

[1]The author is at the European Southern Observatory, Alonso de Cordova 3107, Santiago, Chile, on leave from Service d'astrophysique, CEA-Saclay, France.  E-mail: fmirabel@eso.org

Alternatively, relativistic particles can be injected in the surrounding medium by the wind from a young pulsar (9). In this scenario the slowing rotation of a young pulsar provides stable energy to the non-thermal relativistic particles in the shocked pulsar wind material outflowing from the binary companion (see the figure, right panel). As in the microquasar-jet model proposed for LSI +61 303 by Albert et al., the gamma-ray emission is produced by inverse Compton scattering of the relativistic particles from the pulsar wind on stellar photons. In this context, LSI +61 303 would resemble the gamma-ray binary PSR B1259-63, a radio pulsar in an eccentric orbit around a star of spectral type Be (10).

The compact objects in these three gamma-ray binaries (LS 5039, LSI +61 303 and PSR B1259-63) have eccentric orbits around stars with masses in the range of 10 to 23 solar masses, and these stars provide the seed photons to be scattered by the inverse Compton effect to VHEs. PSR B1259-63 contains a pulsating neutron star, but in LS 5039 and LSI +61 303 the precise nature of the compact stars is not known. Certainly, they are no more than 4 solar masses, which is consistent with neutron stars and/or black holes of low mass. Both, the microquasar-jet model and the pulsar-wind model yield similar mechanisms to produce the VHE emission (that is, the inverse Compton effect). Thus, the fundamental question that remains open is whether the relativistic particles in LS 5039 and LSI +61 303 come from accretion powered jets or from the rotational energy of pulsars that are spinning down as in PSR B1259-63.

The pulsar wind model requires gamma-ray binaries with neutron stars young enough to provide large spin down energies. In fact, as in PSR B1259-63, LS 5039 and LSI +61 303 contain young compact objects. Kinematic studies show that LS 5039 has been shot out from the plane of the Galaxy (11) and LSI +61 303 from a cluster of massive stars (12) sometime during the past 1 million years by supernova explosions produced when the compact objects were formed. Furthermore, it has been proposed that LSI +61 303 is a pulsar-wind source because the time variability and the radio and X-ray spectra resemble those of young pulsars (13). Besides, LSI +61 303 contains a Be star like PSR B1259-63, and the high energy emission in both objects seems to be produced at specific phases of orbital motions of the compact objects around the Be stars. All Be/X-ray binaries known so far contain neutron stars and none is known to host a black hole.

However, the pulsar wind model (13) does not satisfactorily explain the GeV gamma-ray and radio wavelength fluxes observed in LSI +61 303 and LS 5039. In addition, contrary to what would be expected in the radio emission from a young pulsar, the jets in LS 5039 are steady and two sided, and seem to have bulk motions of 0.2 to 0.3 times the speed of light, as do the compact jets in black hole microquasars. Furthermore, in LS 5039 no major radio outbursts are observed similar to those in PSR B1259-63.

The detection of pulsations would be a definitive proof for the pulsar-wind mechanism in gamma-ray binaries. On the other hand, detection of VHE emission from a black hole binary (e.g. Cygnus X-1, V 4641, GX 339-4) would provide definitive observational ground to the microquasar-jet model. Another direct way to distinguish between accretion and rotational powered gamma-ray binaries may be to use radio images with higher sensitivity and angular resolution that would establish clearly whether the high-energy particles that trigger the VHE emission emanate as pulsar winds or as highly collimated microquasar jets.

Gamma-ray binaries are becoming subjects of topical interest in high energy astrophysics, and their study has important implications. As microblazars, they would serve as valuable nearby laboratories to gain insight into the physics of distant blazars. As pulsar-wind gamma-ray binaries they are important because they are the likely precursors of a much larger population of high-mass x-ray binaries in the Milky Way, and may provide clues to understand the early evolution of the enshrouded hard x-ray binaries being discovered from space with satellite telescopes such as the European Space Agency's International Gamma Ray Astrophysics Laboratory (INTEGRAL).

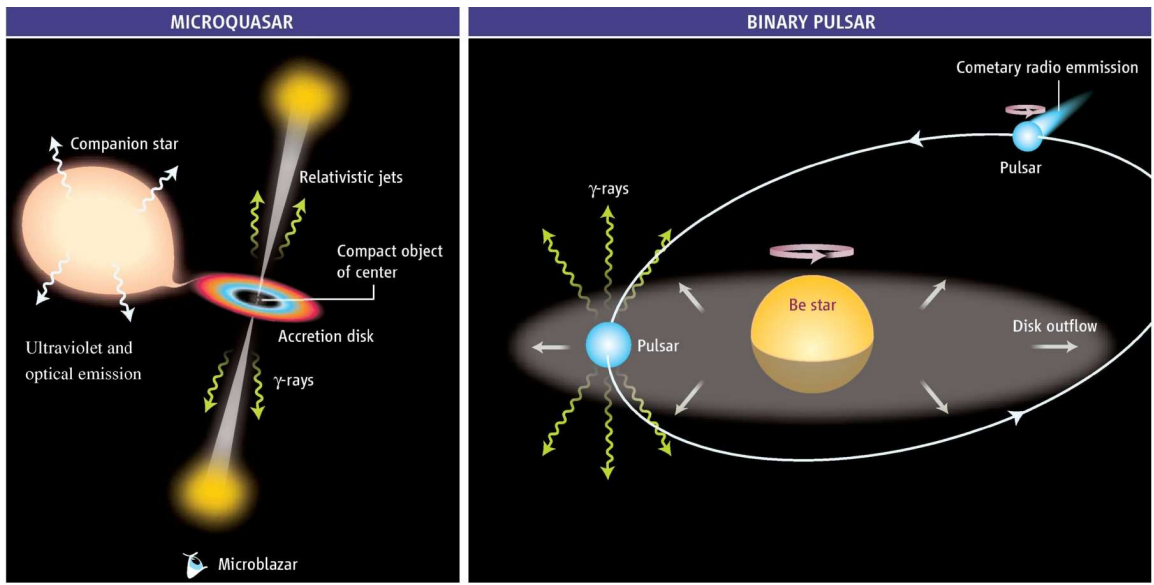

Figure caption: **Alternative models for very energetic gamma-ray binaries**.

**(Left)** Microquasars are powered by compact objects (neutron stars or stellar-mass black holes) via mass accretion from a companion star. This produces collimated jets that, if aligned with our line of sight, appear as microblazars. The jets boost the energy of stellar photons to the range of very energetic gamma-rays.

**(Right)** Pulsar winds are powered by rotation of neutron stars; the wind flows away to large distances in a comet-shape tail. Interaction of this wind with the companion-star outflow may produce very energetic gamma-rays.